\documentclass[letterpaper]{JHEP}
\usepackage{cite}
\newcommand{\fft}[2]{{\frac{#1}{#2}}}

\newcommand{\be}{\begin{equation}}
\newcommand{\ee}{\end{equation}}
\newcommand{\eq}[1]{(\ref{#1})}
\def\nn{\nonumber}
\def\bea{\begin{eqnarray}}
\def\eea{\end{eqnarray}}
\title{Supersymmteric Null-like Holographic Cosmologies}

\author{Feng-Li Lin and Wen-Yu Wen\\
Department of Physics, National Taiwan Normal University\\
Taipei City, Taiwan 116\\
E-mail: \email{linfengli@phy.ntnu.edu.tw},
\email{steve.wen@gmail.com}}

\abstract{We construct a new class of $1/4$-BPS
time dependent domain-wall solutions with null-like metric and dilaton in type II supergravities, which admit a null-like big bang singularity. Based
on the domain-wall/QFT correspondence, these solutions are dual to
$1/4$-supersymmetric quantum field theories living on a boundary
cosmological background with time dependent coupling constant and UV cutoff. In particular we  evaluate the holographic $c$ function for the 2-dimensional dual field theory living on the corresponding null-like cosmology. We find that this $c$ function runs in accordance with the $c$-theorem as the boundary universe evolves,  this means that the number of degrees of freedom is divergent at big bang and suggests the possible resolution of big bang singularity.}

\begin{document}

%%%%%%%%%%%%%%%%%%%%%%%%%%%%%%%%%%%%%%%%%%%%%%%%%%%%%%%%%%%%%%%%%%%%%%%%%%%%
\section{Introduction}
%%%%%%%%%%%%%%%%%%%%%%%%%%%%%%%%%%%%%%%%%%%%%%%%%%%%%%%%%%%%%%%%%%%%%%%%%%%%

One of the important questions in quantum cosmology is to
understand the spacetime structure near big bang or big crunch,
and how the curvature singularity can be resolved by the quantum
effect. String theory is a theory of quantum gravity and is
believed to be able to provide clues for this question. In fact,
some efforts along this direction has been made in the past few
years, see \cite{bb,LMS,HP} and the follow-ups. In these works,
string theory on time dependent supersymmetric backgrounds with
null-like or space-like  singularity are considered, it is hoped
that supersymmetry will diminish the divergence of physical
quantities due to the big-bang, however, it is on the contrary in
the cases considered in \cite{bb,LMS,HP}. Despite that, one would
still hope that some other supersymmetric time dependent
background will help to tame the curvature singularity and to
arrive the viable big bang cosmology.

The UV disaster near the curvature singularity is due to the
divergent gravitational coupling at this region. One may hope that
some strong/weak S-duality will help to understand the strong
gravity behavior by studying its weakly coupled dual theory.
Recently, a null-like time dependent linear dilaton background
preserving $1/2$ supersymmetry is considered in \cite{mbb}, see
also \cite{mbb1}. This background is geodesically incomplete, and
can be served as a toy model studying string theory near the big
bang singularity. Moreover, the simplicity of the background
allows the formulation of the dual matrix theory, which is a
$(1+1)$-d supersymmetric Yang-Mills theory in a time-dependent
world-sheet. The S-duality feature of the string/M-theory
correspondence makes the physics of the singularity under control.

The alternative well-known S-duality happens in the AdS/CFT
correspondence \cite{adscft,gkpw}, where the weak bulk Anti-de Sitter
(AdS) gravity is dual to strongly coupled conformal field theory.
This correspondence is discovered by considering the near horizon
limit of the $D3$-branes, and other similar settings where the
boundary theory is conformal. Later on, it is generalized to so
called domain-wall/QFT correspondence by considering the near
horizon limit of the general $Dp$-branes \cite{imsy,BST}.  In the
domain-wall/QFT correspondence, the bulk background is conformal
AdS space, and the dialton profile is nontrivial, thus, the
boundary theory is no longer conformal. This provides an interesting
setting for the holographic RG running.

Based on the consideration of the S-duality in AdS/CFT (or
domain-wall/QFT) correspondence, we hope to generalize it to the
supersymmetric time dependent background. If the background is
geodesically incomplete, then it provides a new setting to study
the physics of singularity in the context of holography. We will
explore this new direction by constructing the new supersymmetric
time dependent solutions in the context of domain-wall/QFT
correspondence.  We should mentioned that the similar idea in constructing the null-like AdS solutions has also been independently pursued in \cite{sfetsos,ChuHo,das} \footnote{While we were having difficulty in constructing the null-like $AdS_5$ solution based on the ansatz of AdS pp-wave \cite{AdSpp}, we received the draft \cite{ChuHo} by email from P.M. Ho to inform us that they had found the solution. After reading \cite{ChuHo} we realized that we should turn on the time-dependence on dilaton, not on the scalars in \cite{AdSpp}. Then we decide to switch to find the null-like domain-wall solutions discussed in this paper.}.

%Plane wave background with branes and AdS spacetime have been
%studied in various contexts\cite{ppwave}.

In this paper, we construct a new class of $1/4-$BPS domain-wall solution with
null-like dilaton. Our solution is the generalization of the
solution found in \cite{BST} in the context of domain-wall/QFT
correspondence, and it takes the following form in the dual frame (here we omit some prefactors which will be recovered later on.):
\begin{eqnarray}\label{dual1}
&&ds_{dual}^2=r^{-(p-5)}a(u)^2(-2dudv+h(u,r,\vec{x})du^2+d\vec{x}_{(p-1)}^2)+r^{-2}dr^2+d\Omega_{(8-p)}^2\nonumber
\\\nonumber
&&e^{\phi}=r^{-(p-3)(p-7)/4}b(u)\\
&&F_{uv\cdots pr}= (7-p)r^{(6-p)}.
\end{eqnarray}
Here $\phi$ is the dilaton, and $F_{p+2}$ is the Ramond-Ramond
(RR) form flux sourced by $Dp$-branes.  The metrics between string
and dual frames are related by $g_{string}=e^{2\phi\over 7-p}
g_{dual}$. The time dependent profiles $a(u)$ and $b(u)$ will be
determined later on, and will be shown to cause a caustic at
finite proper time interval.

The metric in the dual frame takes
the form of $AdS_{p+2} \times S^{8-p}$, however, the boundary of
the AdS space is now a pp-wave background dressed by a time
dependent Weyl factor. This induces a boundary cosmology.
Our solutions then provide a holographic dual for a QFT
living on $p+1$-dimensional cosmological background with a big
bang-like singularity. In fact, the time profile $a(u)$ is the
scale factor of the boundary cosmology and we will show that it
obeys Friedman-like equation. Our solutions are in some sense the
supersymmetric version of the braneworld \cite{brane} or the
mirage cosmology \cite{mirage} induced by the time-varying bulk
configurations. Furthermore, we use the test probes to derive the the effective gauge coupling and the energy-distance relation. We find that both are time dependent. In this way, our solutions also provide a model to study the quantum field theory with time dependent coupling and UV cutoff.

It is then interesting to use the perspective of the
domain-wall/QFT correspondence to understand the QFT in a
time-dependent background, especially its behavior near the big
bang. One way to examine this is to count the number of degrees of freedom of the dual field theory near the big bang.  In this paper we evaluate the holographic $c$-function for $p=1$ case, which characterizes the running behavior of the number of degrees of
freedom of the 2-dimensional  dual field theory as the boundary universe evolves. Our result suggests the possible resolution of big bang singularity.

This paper is organized as following: In the next section, we will
solve the equations for the ansatz (\ref{dual1}) and determine the
time dependent profiles $a(u)$ and $b(u)$. We then show that the
solution preserves $1/4$ supersymmetry. In section 3, we will show
that the metric is geodescially incomplete, thus it admits a
caustic which play the role of big bang singularity. In section 4,
we will discuss the time dependent coupling constant of the dual
quantum field theory from the point of view of domain-wall/QFT
correspondence. In section 5, we evaluate the holographic c
function and found that it is frame-independent, and runs in
accordance with the $c$-theorem as the boundary universe evolves.
We conclude the paper in section 6 with some discussions. In
Appendix A, we briefly review the domain-wall solutions and their
KK reductions. In Appendix B we give some details of the Killing
spinor equations. In Appendix C, for completeness we reproduce the
solution already found in \cite{ChuHo} for $AdS_5 \times S^5$ case
with null-like axi-dilaton field. In  Appendix D, we record the
new solutions for the $AdS_3\times S^3 \times M_4$ with self-dual
3-form flux and null-like axi-dilaton. The more general solutions
can be obtained by applying S-duality to the solutions in
Appendices C and D.

\bigskip

%%%%%%%%%%%%%%%%%%%%%%%%%%%%%%%%%%%%%%%%%%%%%%%%%%%%%%%%%%%%%%%%%%%%%%%%%%%%%%
\section{Domain-wall solutions in null-like dilaton background}

We consider the ten-dimensional type II supergravity, and the
action of the bosonic sector in Einstein frame is
\begin{equation}
{\cal L}=R-\fft12 (\partial\phi)^2 - \fft{1}{2\cdot
(p+2)!}e^{\sigma\phi} F_{(p+2)}^2.
\end{equation}
The constant characterizing the coupling between dilaton and RR
flux is given by  $\sigma={3-p \over 2}$.

These fields can support soliton-like objects extended along
$(p+1)$ subspace, which are further interpreted as a stack of N
D$p$-branes from the viewpoint of string theory.  Domain-wall
solutions were obtained by taking near-horizon limit\cite{BST}.
Only in the case of $p=3$, taking near-horizon limit explicitly
gives us $AdS_5\times S^5$ bulk geometry and supersymmetry is
enhanced from $1/2-$BPS to maximum\footnote{This enhancement also
happens while taking near-horizon limit of M$2$ and M$5$ branes in
M-theory, due to the lacking of dilaton field in the $D=11$
supergravity.}.  The AdS/CFT correspondence conjectures that the
same physics can also be described in term of ${\cal N}=4, SU$(N)
superconformal field theory living on the boundary of $AdS_5$.  In
the case of generic $p\neq 3$, the bulk is no longer $AdS$
geometry but domain-wall with warp factor supported by
non-vanishing dilaton field.  The radius-dependent dilaton also
breaks conformal symmetry on the boundary, while still preserving
$1/2-$BPS without enhancement.

In this section we will generalize the domain-wall solution ($p\ne
3$) found in \cite{BST} to the ansatz given in (\ref{dual1}) by
introducing the nontrivial time dependent profiles $a(u)$ and
$b(u)$. We will further show that the time-dependence breaks the
supersymmetry to $1/4$.

%%%%%%%%%%%%%%%%%%%%%%%%%%%%%%%%
\subsection{Solving the equations of motion}

The metric in Einstein frame is related to the one in dual frame  by
$g_{einstein}=e^{{p-3\over 2(7-p)}\phi}g_{dual}$, then from
(\ref{dual1}) our ansatz in the Einstein frame becomes
\begin{eqnarray}\nonumber
&&ds^2_{einstein}=r^{(p-7)^2/8}a(u)^2b(u)^{(p-3)/(2(7-p))}(-2dudv+h(u,r,\vec{x})du^2+d\vec{x}^2_{(p-1)})\\
&&\qquad\qquad  +r^{(p-3)^2/8}b(u)^{(p-3)/(2(7-p))}(r^{-2}dr^2+d\Omega^2_{(8-p)})
\end{eqnarray}
and the dilaton and the form flux are the same as in \eq{dual1}.
We will use the indices
$i,j,k$ for the coordinates $\vec{x}$ in the flat metric $d\vec{x}_{(p-1)}^2$, and the indices
$m,n,l$ for the ones in $d\Omega_{(8-p)}^2$.

The equation of motion reads:
\begin{eqnarray}
&&R_{MN}=\fft12\partial_M \phi \partial_N \phi + \fft{e^{\sigma\phi}}{2 (p+1)!}
(F_{MK_1\cdots K_{p+1}}F_N^{\;\;K_1\cdots K_{p+1}}-\fft{p+1}{8(p+2)}g_{MN}F_{(p+2)}^2) \label{eqn:eom_riemann}\\
&&\nabla_M (e^{\sigma\phi} F^{MN_1\cdots N_{p+1}})=0  \label{eqn:eom_bianchi}\\
&&\nabla^2 \phi =\fft{\sigma}{2(p+2)!} e^{\sigma\phi} F_{(p+2)}^2
\label{eqn:eom_dilaton},
\end{eqnarray}
where equations (\ref{eqn:eom_bianchi}) and (\ref{eqn:eom_dilaton})
are satisfied provided that
\be\label{cons1}
 b(u)=(a(u))^{\fft{(p+1)(p-7)}{2(p-3)}}.
\ee

It is then straightforward to calculate
\begin{eqnarray}
&&F_{(p+2)}^2=-(p+2)!(7-p)^2 r^{-(p-6)(p-3)^2/8}a^{(p+1)(p-6)/4}\nonumber\\
&&F_{MK_1\cdots K_{p+1}}F_N^{\;\;K_1\cdots K_{p+1}}=\fft{1}{p+2}g_{MN}F_{(p+2)}^2 \nonumber\\
&&\sqrt{-g}e^{\sigma\phi}F^{uv\cdots p r}=p-7
\end{eqnarray}
and \eq{eqn:eom_riemann} are solved by given Ricci tensors,
\begin{eqnarray}
&&R_{uv}= \fft{(7-p)^3}{16}r^{(5-p)}a^2\\
&&\label{ruu} R_{uu}=
\fft{(p-7)^3}{16}r^{(5-p)}a^2 h+\fft{p^2-6p-23}{8}(\partial_u\ln a)^2+2\partial^2_u \ln a-\fft12\vec{\nabla}^2h \nonumber\\
&&+\fft12 r^{(5-p)}a^2 [(p-8)r\partial_rh-r^2\partial_r^2h]\\
&&R_{ij}= -\fft{(7-p)^3}{16}r^{(5-p)}a^2\delta_{ij}\\
&&R_{rr}= \fft{(p+1)(p-5)(7-p)^2}{32}r^{-2}\\
&&R_{mn}=\fft{(p+1)(7-p)^2}{16}g_{mn}^{((8-p)-sphere)}
\end{eqnarray}
where $\vec{\nabla}^2$ is the $(p-1)$-dimensional Laplacian and
$g^{((8-p)-sphere)}_{mn}$ is the metric of the unit
$(8-p)$-sphere.

While others are trivially satisfied, only the $uu$-component of
 \eq{eqn:eom_riemann} gives nontrivial constraint on functions
 $a$ and $h$, that is,
\begin{eqnarray}\label{timep}
2\partial_u^2 \ln a-\fft{32}{(p-3)^2} (\partial_u\ln a)^2
= \fft12 \vec{\nabla}^2 h -\fft12
r^{(5-p)}a^2[(p-8)r\partial_rh-r^2\partial_r^2h].
\end{eqnarray}

  We should remind the reader, the above solutions are for  $p\ne 3$ cases. For $p=3$, we need to impose the self-dual condition on the 5-form flux. The solution has been found in \cite{ChuHo}, for completeness we include it in the Appendix C, and the analogue of \eq{timep} is given in \eq{ads5_eom_uu}.

\subsection{Supersymmetry analysis}

To check how much supersymmetry preserved by the solution found
above, we should look into the Killing spinor equations
\cite{Berg}
\begin{eqnarray}
&&\delta \Psi_M = \partial_M\epsilon -\fft14
\omega_M{}^{ab}\gamma_{ab}\epsilon + \fft{(-)^p}{8(p+2)!}e^{\phi}F\cdot\gamma \gamma_M\epsilon^{\prime}=0\label{susy_gravitino}\\
&&\delta \lambda = \gamma^M\partial_M\phi \epsilon +
\fft{3-p}{4(p+2)!}e^{\phi}F\cdot \gamma
\epsilon^{\prime}=0\label{susy_dilatino}
\end{eqnarray}
where $\Psi_M$ and $\lambda$ are gravitino and dilatino respectively. These are the Killing spinor equations in the string frame. To examine these equations, we need to transform the metric in \eq{dual1} into string frame and derive the corresponding spin connections. The details are given in the Appendix B.

Following the metric ansatz in the string frame
\eq{eqn:metric_string}, the variation of dilatino
\eq{susy_dilatino} reads
\begin{eqnarray}
\delta \lambda &=& \gamma^u\partial_u\phi \epsilon +
\gamma^r\partial_r\phi + \fft{3-p}{4 (p+2)!}e^\phi F\cdot \gamma
\epsilon^{\prime} \nonumber\\
&=& \fft{\partial_u b }{b}\gamma^u\epsilon +
\fft{-1}{4r}(p-3)(p-7)\gamma^r
(\epsilon+(-)^p\bar{\gamma}^{+-\cdots p}\epsilon^{\prime})=0.
\end{eqnarray}
Here we have used constraint \eq{cons1} obtained from solving
equations of motion. This Killing equation vanishes if two
projections are imposed on an arbitrary constant spinor,
\begin{equation}
\gamma^u\epsilon_0=0, \qquad
\epsilon_0+(-)^p\bar{\gamma}^{+-\cdots p}\epsilon_0^{\prime}=0
\end{equation}
Given the spin connections in \eq{eqn:connection}, variation of
gravitino \eq{susy_gravitino} is given by
\begin{eqnarray}
\delta \Psi_+&=& \partial_+\epsilon - \fft14
r^{\alpha}a^{-\delta}\partial_rh \bar{\gamma}^{+r}\epsilon -
\fft14 r^{-\alpha}a^{-\beta}\partial_ih \bar{\gamma}^{+i}\epsilon
\nn\\
&-&\fft12 \alpha r^{\alpha-1}a^{-\delta}\bar{\gamma}^{-r}(\epsilon+(-)^p\bar{\gamma}^{+-\cdots p}\epsilon^{\prime})\label{eqn:gravitino+}\\
\delta \Psi_-&=& \partial_-\epsilon - \fft12 \alpha
r^{\alpha-1}a^{-\delta}\bar{\gamma}^{+r} -\beta r^{-\alpha}
a^{-\beta-1}\partial_u a (\bar{\gamma}^{+})^2
\nn\\
&+& \fft12 (-)^p \alpha r^{\alpha-1} a^{\delta} \bar{\gamma}^{+-\cdots pr}\bar{\gamma}_-\epsilon^{\prime}\label{eqn:gravitino-}\\
\delta \Psi_i&=& \partial_i\epsilon - \beta r^{-\alpha}
a^{-\beta-1}\partial_u a \bar{\gamma}^{i+}\epsilon
-\fft12 \alpha r^{\alpha-1} a^{-\delta} \bar{\gamma}^{ir}(\epsilon+(-)^p\bar{\gamma}^{+-\cdots p}\epsilon^{\prime})\label{eqn:gravitinoi}\\
\delta \Psi_r&=& \partial_r \epsilon - \delta
r^{-\alpha}a^{-\beta-1}\partial_u a\bar{\gamma}^{r+}\epsilon
+ \fft12 (-)^p \alpha r^{\alpha-1} a^{-\delta} \bar{\gamma}^{+-\cdots p}\epsilon^{\prime}\label{eqn:gravitinor}\\
\delta \Psi_m&=&\partial_m\epsilon -\delta
r^{-\alpha}a^{-\beta-1}\partial_u a \bar{\gamma}^{m+}\epsilon
-\fft12\alpha
r^{\alpha-1}a^{-\delta}\bar{\gamma}^{mr}(\epsilon+(-)^p\bar{\gamma}^{+-\cdots
p}\epsilon^{\prime})\nonumber\\
&+&\fft12(2\alpha-1)r^{\alpha-1}a^{-\delta}\bar{\gamma}^{mr}\epsilon\label{eqn:gravitinoa},
\end{eqnarray}
where
$\alpha=\fft{7-p}{4},\beta=\fft{p-7}{2(p-3)},\delta=-\fft{p+1}{2(p-3)}$.
Equations \eq{eqn:gravitino+},\eq{eqn:gravitino-} and
\eq{eqn:gravitinoi} are solved by imposing same projections as in
transformation of dilatino, the remaining \eq{eqn:gravitinor} and
\eq{eqn:gravitinoa} are solved by additionally letting Killing
spinors have coordinates dependence, that is
\begin{equation}
\epsilon=e^{\fft12
r^{\alpha}a^{-\delta}}e^{\fft12(1-2\alpha)r^{\alpha-1}a^{-\delta}\bar{\gamma}^{mr}\Omega^m}\epsilon_0
\end{equation}
Thus, we conclude our solutions preserve $1/4$ supersymmetry.

\section{Boundary cosmologies and geodesic incompleteness}

In this section we will discuss some properties of the solutions
found above. Especially, we will make the dynamics of the boundary
cosmology explicit by solving the scale factor $a(u)$.

A related issue is about the choice of the frames when we discuss
the boundary cosmology. Although we solve the field equations in
the Einstein frame, it would be more natural to discuss the
holographic boundary cosmology in the dual frame since the bulk
metric is $AdS_{p+2}\times S^{8-p}$ and the usual arguments for
AdS/CFT correspondence  can be generalized easily \cite{BST}.
Moreover, one can KK reduce the 10-dimensional supergravity to
$(p+2)$-dimensional one because there is no warped factor in front
of the $S^{8-p}$ metric. In the following, we will stick to the
dual frame metric when we discuss the dynamics of the boundary
cosmology.

\subsection{Boundary cosmologies}
We are now ready to solve \eq{timep} for the scale factor
$a(u)$ and the pp-wave-front profile $h(u,r,\vec{x})$ of the boundary cosmology.

Since the L.H.S. of \eq{timep} is function of u only, we can solve
the function $h$ by \be h(u,r,\vec{x})= {2 P(u) \over
(p-5)}r^{(p-5)}+h_0(u,\vec{x})r^{(p-7)}+h_1(u,\vec{x}) \ee with \be
\vec{\nabla}^2 h_0(u,\vec{x})=0, \qquad \vec{\nabla}^2 h_1(u,\vec{x})=4 Q(u)
\ee
and $P(u)$ and $Q(u)$ are arbitrary functions of $u$.

Plugging this into \eq{timep}, we get the equation for
$a(u)$ as following \be\label{Fpro}
\partial_u^2 \ln a -\kappa (\partial_u \ln a)^2=Q(u)+P(u)a^2.
\ee
where $\kappa\equiv 16/(p-3)^2$.

On the other hand, we can make the dynamics of the boundary
cosmology more explicit by rewriting \eq{Fpro} into the form of
the Friedman equation. To do this we first introduce the time
coordinate $t$ which is related to the conformal time $u$ by
\be
dt\equiv a(u) du,
\ee
The equation \eq{Fpro} can be put into the form of Friedman
equation
\be\label{Frie} {\ddot{a} \over a} - \kappa H^2= {Q(u)
\over a^2}+P(u)
\ee
where the dot denotes the derivative with
respect to $t$, and the Hubble parameter is defined as
\be
H={\dot{a}\over a}\;.
\ee

Since \eq{Fpro} is a nonlinear differential equation, it is hard
to solve the general solution for generic $P(u)$ and $Q(u)$,
however, once these functions are given, one can solve the scale
factor to yield various kind of cosmologies. In the following, we
will solve the scale factor in some simple cases:

(i) The simplest case is to set $P=Q=0$, then the solution is
\be
 a(u)=(c_0+c_1u)^{-1\over \kappa}
\ee where $c_{0,1}$ are
arbitrary constants. Obviously the scale factor $a(u)$ is singular
if $c_0+c_1u=0$.

(ii) The next case is $P=0$ but $Q$ is a constant, then the
solution is \be a^2(u)=c_0 \cos(\sqrt{\kappa Q}u)+c_1
\sin(\sqrt{\kappa Q}u), \quad Q>0, \ee or \be a^2(u)=c_0
\exp({\sqrt{\kappa |Q|}u})+c_1 \exp(-\sqrt{\kappa |Q|}u), \quad
Q<0. \ee The $Q>0$ case is the type of closed universes, and the
$Q<0$ case is the type of open universes. From (i) and (ii), it
seems that $Q$ plays the similar role of the spatial curvature
constant.

(iii) The last case is $Q=0$ but $P$ is a constant, and the
solution is \be \pm u +c_1 =\int^{F=a^{-\kappa}} {dy \over
\sqrt{c_0-{\kappa^2 P\over \kappa-1}y^{2-{2\over \kappa}}}}\;. \ee
The solution will be some hypergeometric function.

Finally, we would like to mention that \eq{Frie} is different from
the Friedman equation derived from the Einstein gravity based on
the null-like boundary metric:
$ds^2=a^2(-2dudv+h(u)du^2+dx_{p-1}^2)$. For such a metric, only
the $uu$-component of the Einstein tensor is nonzero, and the
Einstein gravity yields the following Friedman equation \be
{\ddot{a}\over a} -2 H^2={8\pi G_N \over p-1}\rho_m+{2-p\over 2}h.
\ee Here $\rho_m$ is matter's energy density.

In our case, the Friedman-like equation \eq{Frie} should be the
one derived from the effective gravity on the boundary induced by
the bulk geometry, which may not be the Einstein gravity due to
the nontrivial counter terms \cite{CL}. Moreover, from the dual
theory point of view, these counter terms come from the dual QFT's
contributions. It will be interesting to understanding the origin
of the functions $P(u)$ and $Q(u)$ from the dual QFT.

\subsection{Geodesic incompleteness}

Our solutions are the vacuum solutions of type II supergravity, so
the curvature invariants will be constant  finite value determined
by the strength of the RR-flux as for the Freund-Rubin cases.
However,  for some cases the scale factor solved in last subsection is
singular as the conformal time $u$ approaches some finite value.
We will see that the singularity leads to some caustic which
terminates the geodesics. This shows that our solutions contain the
null-like big bang singularity.

It is easy to see that the $uu$-component of the Ricci tensor,
i.e., \eq{ruu} is also singular if the scale factor is singular.
This will lead to the caustics by the Raychaudhuri equation
\cite{HE} \be {d\theta \over d s}=-R_{MN} V^M V^N+
(V^M{}_{ ;N}V^N)_{;M} \ee where $V=V^u \partial_u$ is the unit
velocity vector field for a congruence of the null-like curves,
and $\theta$ is the expansion parameter of the volume enclosed by
the congruence.  Therefore, as $R_{uu}$ becomes singular, the
expansion parameter will shrink to zero at finite value of the
affine parameter $s$ and cause the caustics.

Moreover, we would like to make a connection between the affine
parameter $s$ and the conformal time coordinate $u$, and see if
the caustics happen at finite $s$ or not. In fact, we are just generalizing
the treatment in \cite{mbb} straightforwardly to our case, see also \cite{mbb1,ChuHo}. This is done by the
geodesic equation
\be {d^2 u \over d s^2}+\Gamma^u_{u u}({d u
\over ds})^2=0
\ee
where
\be
\Gamma^u_{uu}=2 \partial_u \ln a
\ee
for both the bulk and boundary metrics. From the geodesic
equation it is easy to see that
\be
{du\over ds}={1\over a^2}\ge 0.
\ee
This shows that $s$ is a
monotonic function of $u$, and implies that the geodesics
terminate at finite value of $s$ if the singularity of the scale
factor occurs at finite value of $u$.

Finally, we would like to mention, not all of the solutions
contain caustics. For example, for the case (ii) with $Q<0$, $c_0>
0$ and $c_1\ge 0$ in the previous subsection, the scale factor
never shrinks to zero at finite $u$.

\section{Probing the dual quantum field theory}

In this section we would like to study the dual quantum field
theory from the point of view of test probes.  As pointed out in
\cite{PP}, there are two different probes: the
closed string one and the open string one \footnote{In \cite{PP}, the closed string probe is named as holographic/supergravity probe, and the open string one as the D$p$-brane probe. }. They have different energy-distance relations \cite{UVIR} which relates the the UV cutoff of dual field theory to the supergravity radial coordinate.  Our setting
generalizes the ones studied in \cite{imsy, BST, PP} by having an
additional dynamical scale, which is the boundary Hubble scale.

Since the dual quantum field theory is a $p+1$-dimensional
$SU(N)$ Yang-Mills theory, from the dimensional analysis, the effective
dimensionless gauge coupling is related to the dimensionful Yang-Mills
coupling by 
\begin{equation}\label{geff}
g_{eff}^2\sim  g_{YM}^2 N E^{p-3}
\end{equation}
where $g_{YM}$ is the UV bare Yang-Mills coupling and $E$ is the UV cut-off for the dual field theory. As will be shown that $g_{YM}$ and $E$ are time dependent. Then we want to express the above relation in terms of supergravity variables for either closed string or open string probes, especially to determine probes' energy-distance relations.

To carry this out, we render our metric ansatz
\eq{dual1} in the dual frame with proper prefactors:
\begin{eqnarray}\label{dualm}
&&ds_{dual}^2=\alpha'[(\bar{g}^2_{YM}N)^{-1}r^{5-p}a(u)^2 ds^2_{(p+1)}+r^{-2}dr^2+d\Omega^2_{(8-p)}]\\
&&\label{gt} e^{\phi}=\fft{1}{N}[(\bar{g}_{YM}^2N)r^{p-3}]^{(7-p)/4}b(u)
%\\&&F_{uv\cdots r}=(7-p)\alpha'\lambda^{-1}r^{6-p}
\end{eqnarray}
where $\bar{g}_{YM}^2:=g_s (\alpha')^{(p-3)/2}$ with $g_s$ being defined in \eq{gs} is the UV bare Yang-Mills coupling in the static domain-wall/QFT correspondence. The string frame metric is Weyl-related by $g_{st}=(Ne^{\phi})^{2/(7-p)}g_{dual}$.

  The simplest closed string probe is the dilaton which couples to the boundary gauge invariant operator in string frame as following $e^{-\phi} \sqrt{-g} F^2$. This implies that the dimensionless effective gauge coupling for closed string probe in string frame is
\be\label{geff1}
g^2_{eff} \sim Ne^{\phi}=[g^2_{YM} Nr^{p-3}]^{7-p\over 4}
\ee
where we define the time dependent UV bare Yang-Mills coupling in the above and in \eq{geff} by
\be\label{gym1}
g_{YM}:=\bar{g}_{YM} (b(u))^{2/(7-p)}.
\ee
This is also true for $p=3$ case; for $p\ne 3$ we can further use \eq{cons1} to convert $b(u)$ into $a(u)$.

  We can read off the UV cutoff $E$ by comparing \eq{geff} and \eq{geff1}, and the result is
\be
E\sim (Ne^{\phi})^{1/(7-p)} \fft{r^{(5-p)/2}}{g_{YM}N^{1/2}}
\ee
This is the energy-distance relation for the closed string probe in string frame.   Remarkably, this energy scale is nothing but $\sqrt{g_{tt}/\alpha'}/a(u)$ in string frame except that we should replace $\bar{g}_{YM}$ by $g_{YM}$. Note that $a(u)$ is the scale factor of the boundary cosmology  \footnote{The replacement of $\bar{g}_{YM}$ by $g_{YM}$ may suggest that the warped factor involved $\bar{g}^2_{YM}N$ in \eq{dualm} should be replaced by $g^2_{YM}N$. If so, then the scale factor $a(u)$ of the boundary metric will be changed to $(a(u))^{4/(3-p)}$. For $p<3$, the qualitative behavior of boundary cosmology does not change.} governed by \eq{Frie}.

  On the other hand, the energy-distance relation for the open string probe in string frame is
\begin{equation}\label{openuvir}
E\sim  r.
\end{equation}
This is obtained by considering a string stretched from the origin
to a test brane such that the energy is proportional to its
length. Note that when evaluating $E$ in string frame, the nontrivial warped factor in $\sqrt{g_{rr}}$ is cancelled out by the one in $\sqrt{g_{tt}}$  \cite{BST,PP}. Then from \eq{geff},  \eq{gym1} and \eq{openuvir} we obtain the effective gauge coupling for the open string probe
\be
g^2_{eff} \sim g_{YM}^2 N r^{p-3}.
\ee

 We can transform the gauge invariant operator $\sqrt{-g} F^2$ in string frame to the one in dual frame and absorb the Weyl factor into effective gauge coupling. In this way we can obtain the effective gauge coupling and energy-distance relations measured in dual frame. The results are
\be \label{uvirg}
g^2_{eff}\sim [g_{YM}^2 Nr^{p-3}]^{(5-p)/2}
\ee
and
\be \label{uvird}
E \sim
\fft{r^{(5-p)/2}}{g_{YM}N^{1/2}}
\ee
for both the closed and open string probes \footnote{The relations \eq{uvirg} and \eq{uvird} are the generalization of the ones for the closed string probe in \cite{PP} with $\bar{g}_{YM}$ replaced by $g_{YM}$.}. Note that the energy-distance relation \eq{uvird} can be either understood as the energy of a string with length $r$, namely, $E\sim \sqrt{g_{rr}g_{tt}} r\sim (Ne^{\phi})^{2/(p-7)}r $ for open string probe, or as $\sqrt{g_{tt}/\alpha'}/a(u)$ but with $\bar{g}_{YM}$ replaced by $g_{YM}$ for closed string probe.

\bigskip

  Few remarks are in order for our above result.
\begin{itemize}
\item Although we can obtain effective gauge coupling and energy-distance relation in different frame by Weyl transformation induced by some power of $(Ne^{\phi})$, it is more natural to work in the dual frame since its radial metric is time independent. On the contrary, the time dependence of the Weyl factor will dress the radial metric in the other frames.

  Moreover, the effective gauge coupling in the dual frame is the same for both closed and open string probes as in the $p=3$ cases.

\item   The dynamical quantities $g_{YM}$, $g_{eff}$ and $E$ are all time dependent.  Especially, at the big bang the couplings are vanishing and $E$ is divergent for $p<3$ cases if $r$ is fixed. Therefore, the $p<3$ dual field theory is valid effectively near the big bang with vanishing coupling.

\item For the boundary effective field theory description to be valid, we should require
\be
E>H
\ee
where $H:=\dot{a}/a$ is the Hubble scale derived from the boundary cosmology, which plays the role of a dynamical energy scale for the dual field theory.

This is the additional
constraint besides the usual ones in the static domain-wall/QFT
correspondence, namely, (i) $g^2_{eff}\ll 1$ for the validity of
the perturbation of the dual field theory and (ii) small curvature
($g^2_{eff}\gg 1$) and small dilaton ($e^{\phi}\ll 1$) for the
validity of classical supergravity theory. Note that these
conditions are now time dependent.  Especially, at big bang,  supergravity is invalid because of small effective coupling (large curvature) but the dual description is of no problem for $p<3$ cases.

\end{itemize}

\section{Cosmic c functions from holography}

An interesting question for the big bang cosmology is the
evolution of the number of degrees of freedom, especially, we
would like to know the amount of degrees of freedom near the bang
bang singularity. Intuitively this is related to the issue of
resolving the space-like singularity. For example, in the conifold
transition \cite{conifold}, the singularity is resolved with the
emergence of light degrees of freedom by wrapping the tensionless
D-brane.

For a conformal field theory, the number of degrees of freedom is
characterized by the central charge which can be extracted from
the coefficient of Weyl anomaly. For non-conformal field theory,
the central charge is no longer constant, which is instead called
the $c$ function and will run with the energy scale. Moreover, a
$c$-theorem for 2-diemsnional field theory is proved in \cite{zz}
that the $c$ function will never increase in a RG flow. However,
for higher dimensional cases, there is no rigorous proof of
$c$-theorem as far as we know.

In this section we would like to evaluate the c function of the
dual field theory from the bulk gravity for our null-like
cosmological background. Usually one can extract the $c$ function
from the Weyl anomaly which exists only in even dimensional
space-time. For simplicity, we consider the $p=1$ case which
yields a 2-dimensional field theory living on a time-dependent
background. It turns out the $c$ function is time dependent and
runs in a manner in accordance with $c$-theorem as the universe
evolves, thus the cosmological evolution induces the RG flow for
the dual field theory. Moreover, it leads to a fact that the
number of degrees of freedom is divergent when approaching the big
bang, and suggests the possible resolution of the big bang
singularity.

  It is also interesting to evaluate the $c$ function for $p=5$ and examine the singularity issue, however, in this case the effective gauge coupling diverges and the UV cutoff is vanishing so that the validity of the dual field theory is in question.

  Before starting the calculations, we would like to comment on the holographic $c$ function for the solutions recorded in Appendix C and D. In these cases, either the dilaton is a constant or it does not couple to the form-flux, therefore there is no non-trivial scalar potential so that the $c$ function runs with neither $r$ nor $u$, i.e. these are still central charges.

\subsection{Holographic $c$ function}

The canonical method in evaluating the RG flow from holography was
developed in \cite{BVV} by applying the Hamilton-Jacobi
formulation of the bulk gravity to construct the counter terms of
the bulk gravity action. In this formalism the Hamilton-Jacobi
functional is interpreted as the quantum effective action of the
dual field theory resulting from integrating out the matter
degrees of freedom coupled to the boundary gravitational
background.  Moreover, though the field equations of bulk gravity
is of second order, the evolution equation of the quantum
effective action derived from the Hamilton-Jacobi formalism, is of
first order and takes the form of the Callan-Symanzik equation.
This is the RG equation for the dual field theory derived from
holography.

For concreteness we focus on $p=1$ case, and we decompose the
metric into \be\label{adm1}
ds^2=d\rho^2+\gamma_{\mu\nu}(\rho,x^{\mu}) dx^{\mu} dx^{\nu}, \ee
where $\rho$ is the radial coordinate, and $\gamma_{\mu\nu}$ is
the metric for the 2-dimensional transverse hypersurface with
coordinates $x^{\mu}=u,v$.

The bulk gravity action can be obtained from \eq{draction} by
setting $p=1$ and $\phi=\ln \Phi$, and it is \be S_{bulk}=\int
d^3x \sqrt{-g} \left( R+{1\over 2} G(\Phi) (\partial \Phi)^2 +
V(\Phi) \right):=\int d^3x \sqrt{-g}  {\cal L} \ee with \be
G(\Phi)=-{16\over 9} \Phi^{-2}, \qquad V(\Phi)={1\over 2}
\Phi^{4\over3}. \ee Moreover, its field equations can be solved by
the null-like $p=1$ domain-wall solution \bea
&& ds_3^2=\Phi^{-4/3}[r^4a(u)^2(-2dudv+h(u,r)du^2)+r^{-2}dr^2]\\
&& \Phi=r^{-3} a(u)^3\label{phif}
\eea
where the functions $a(u)$ and $h(u,r)$ are given in section 3.

Following the usual ADM formalism with respect to the metric ansatz \eq{adm1}, one can get the super-hamiltonian constraint as  following
\be\label{hamc}
{\cal H}:=\pi^2 -\pi_{\mu\nu} \pi^{\mu\nu}+{1\over 2 G} \Pi^2 -{\cal L}=0
\ee
where  $\pi_{\mu\nu}$ and $\Phi$ are the canonical ADM momenta for the metric $\gamma_{\mu\nu}$ and the scalar $\Phi$ respectively, and $\pi=\gamma^{\mu\nu}\pi_{\mu\nu}$. From their defining equations, one can obtain the following ``flow equations"
\be\label{flows}
\gamma'_{\mu\nu}=2(\pi_{\mu\nu} - \gamma_{\mu\nu} \pi), \qquad \Phi'=G^{-1} \Pi
\ee
where the prime $'$ denotes the derivative with respect to $\rho$.

In the Hamilton-Jocobi theory, the canonical ADM momenta are
related to the Hamitlon-Jacobi functional $S$ as following
\be\label{canp} \pi_{\mu\nu}={1\over \sqrt{-\gamma}} {\delta S
\over \delta \gamma^{\mu\nu}}, \qquad \Pi={1\over \sqrt{-\gamma}}
{\delta S \over \delta \Phi}. \ee
On the other hand, in the
formalism of \cite{BVV} the Hamilton-Jacobi functional is
interpreted as the quantum effective action of the dual field
theory, which usually contains the local renormalized part and the
non-local part as following
\be\label{seff} S=\int dx^2
\sqrt{-\gamma} \left( Z(\Phi) R+ {1\over 2} M(\Phi) (\partial
\Phi)^2+ U(\Phi)\right) + \Gamma[\gamma_{\mu\nu},\Phi,
\partial_{\mu}^{-1}].
\ee
Note that the vacuum expectation values of the boundary stress
tensor and the boundary gauge-invariant operator $O_{\Phi}$ to
which $\Phi$ couples are \be \langle T_{\mu\nu} \rangle:={1\over
\sqrt{-\gamma}} {\delta \Gamma \over \delta \gamma^{\mu\nu}},
\qquad \langle O_{\Phi} \rangle :={1\over \sqrt{-\gamma}}{\delta
\Gamma \over \delta \Phi} \ee and the Weyl's anomaly is given by
\be\label{cf} \gamma^{\mu\nu} \langle T_{\mu\nu} \rangle:= \langle
T \rangle=-{c\over 12} R +{\beta\over 2} <O_{\Phi}> \ee where $c$
is the $c$ function charactering the number of effective degrees
of freedom, and $\beta$ is the beta function charactering the RG
running of $\Phi$.

Now we use \eq{seff} to evaluate the canonical momenta given in
\eq{canp}, and then plug the results into the super-hamiltonian
constraint to get the Hamilton-Jacobi equations. Collecting terms
in the Hamilton-Jacobi equations we have the following results:

(1) Putting the potential terms together gives \be
U^2+G^{-1}U'^2=2V \ee which can be solved for the boundary scalar
potential \be U(\Phi)={2\over \sqrt{3}} \Phi^{2/3}. \ee (2)
Comparing scalar's kinetic terms $(\partial \Phi)^2$ and
$\partial^2 \Phi$ has \be UZ''+G^{-1}M'U'+G=0, \qquad UZ'+G^{-1}M
U'=0 \ee which can be solved by \be M={64 \over 9\sqrt{3}}
\Phi^{-8/3}, \qquad Z'={8\over 3\sqrt{3}} \Phi^{-5/3}. \ee (3)
Finally, collecting the linear curvature term(including the  Weyl's
anomaly term) arrives \bea\label{tf}
\langle T\rangle&=& U^{-1} (G^{-1}Z'U'-1) R+  U^{-1} G^{-1} U' \langle O_{\Phi}\rangle \\
&=&-{5 \Phi^{-2/3} \over 2\sqrt{3}}R-{3\over 8} \Phi  \langle
O_{\Phi}\rangle. \eea In the second equality we have used the
solution for $U$ and $Z'$. From \eq{tf}, \eq{cf} and \eq{phif} we
can read off the $c$ and $\beta$ functions for the null-like
background \eq{phif} and the result is\footnote{The $\beta$
function solved here is the same as the one obtained by solving
the flow equations \eq{flows} in the potential dominant limit
\cite{BVV}.} \be c(u,r)= 10 \sqrt{3} \left({r\over a(u)}\right)^2,
\qquad \beta(u,r)=-{3\over 4} \left({a(u)\over r}\right)^3. \ee
This suggests that the $c$ function also runs as the boundary
universe evolves, and the inverse scale factor plays the role of
energy scale so that the $c$ function blows up while approaching
the big bang, i.e. as in the UV limit. This is consistent with $c$-theorem as expected
since the dual field theory is asymptotically free as seen from
negative $\beta$ function or as discussed in section 4, where the
coupling constant is time dependent and becomes weakly coupled as
$a(u)$ decreases.

  The behavior of the $c$ function is quite different from the one in the dS/CFT correspondence \cite{dscf}. In dS/CFT the RG flow is the inverse of the cosmological flow so that the c-theorem holds, and we also have the cosmological horizon to justify the choice. In our case, we do not have the cosmological horizon, and the c-theorem is the only check.

\subsection{Frame (in)dependence}
In the above we have calculated the c function in the Einstein
frame, it is natural to ask what is the c function calculated in
the dual frame which has a simpler metric but with a more
complicated bulk gravity action. In \cite{LinWu} it was argued by
explicit example that the $c$ and $\beta$ functions transform as
the vectors on the $\Phi$-space. That is, if two frames are
related by \be d\tilde{s}^2=\Phi^{2\xi} \left(d\rho^2+ s(\rho)^2
\hat{\gamma}_{\mu\nu}(x^{\mu})dx^{\mu}dx^{\nu}\right):=\Phi^{2\xi}d\rho^2
+ \tilde{s}^2(\rho)\hat{\gamma}_{\mu\nu}(x^{\mu})dx^{\mu}dx^{\nu},
\ee then the ``vectors" on $\Phi$-space transform as \be
\tilde{\beta}:=\tilde{s} {d\Phi \over d\tilde{s}} =\Omega\; s
{d\Phi \over ds}:=\Omega \beta, \qquad \tilde{c}=\Omega\; c \ee so
that the transformation function is defined by \be \Omega^{-1}
:={s \over \tilde{s}} {d\tilde{s} \over ds}=1+\xi \Phi^{-1}\beta.
\ee

In the case of relating dual frame to Einstein frame as given by
\eq{dtoe}, we have $\xi=2/3$, i.e. $\xi=2(p-3)/(p(p-7))$ for
$p=1$, and $\beta=-3\Phi/4$ as given above so that $\Omega=2$.
This shows that the $\beta$ and $c$ functions in the dual frame
are just twice as large as the ones in the Einstein frame.

\section{Discussions and Conclusions}
In this paper, we construct a whole class of the quarter BPS
null-like domain-wall solutions generalizing the domain-wall/QFT
correspondence to the cosmological context. All the solutions we
found are geodesic incomplete, moreover, the boundary metrics are
time dependent and obey the Friedman-like equation. It is
interesting to explore more on the possible behaviors of these toy
cosmologies.

To exploit the power of domain-wall/QFT  correspondence, we first
show that the coupling constant of the dual field theory is time
dependent, and discuss the validity of each effective theory in
different regimes. Moreover, we calculate the holographic $c$
function for some of these new solutions corresponding to
asymptotically free QFT. We find that the $c$ function is also
time dependent and runs in agreement with $c$-theorem as the
universe evolves. We did not check the $c$-theorem for all the
solutions, especially for those non-asymptotically free ones. It
will be interesting to investigate in the future.

From the $c$ function, we know that the number of degrees of
freedom is huge near the big bang either for the null-like
$AdS_5\times S^5$ case with finite constant central charge
or for our $p=1$ solution with divergent $c$ function. This may suggest
that the big bang singularity could be resolved. However, one
needs a more direct way to study this issue by exploiting the
S-duality in the usual AdS/CFT correspondence. It will be
interesting to evaluate the correlation functions of the dual
field theory by the way of bulk-to-boundary propagators
\cite{gkpw}, and see if they can tell us some new physics about
the resolution of the singularity or not.   This was in fact the
concern of the original matrix big bang proposal \cite{mbb}.

We hope that our new solutions will provide a new playground for
studying the strongly coupled QFT in the cosmological background
with time dependent coupling constants. We also hope that the
holographic principle will help us in understanding the issues
such as the resolution of space-like singularity.
%%%%%%%%%%%%%%%%%%%%%%%%%%%%%%%%%%%%%%%%%%%%%%%%%%%%%%%%%%%%%%%%%%%%%%%%%%%%
\bigskip
\acknowledgments
We would like to thank Chong-Sun Chu and Pei-Ming Ho for discussions, especially for  showing us their draft \cite{ChuHo} before they submit to arXiv. FLL thanks Miao Li for useful discussions, and also thanks VSOP for their hospitality 
in holding a workshop PLSS-06 "Physics for Large and Small Scales "  
at Hanoi for him to present  this work. This work is supported by Taiwan's NSC grant 94-2112-M-003-014.

\begin{appendix}

\section{Domain-wall solutions and dimensional reductions}
In this section, we briefly review the D$p$-brane solution and
its near-horizon domain-wall geometry, while introducing the
transformation between the Einstein, string and dual frame.

  The solution for $Dp$-brane in the string frame is given by
\begin{eqnarray}
&&ds_{string}^2=
H^{-1/2}d\vec{x}^2_{(p+1)}+H^{1/2}(dr^2+r^2d\Omega^2_{(8-p)}), \\
&&e^\phi=g_s H^{-(p-3)/4}, \label{gs} \\
&&F_{0\cdots pr}=\partial_r H^{-1},
\end{eqnarray}
where $H=1+ \bar{g}_{YM}^2 N
\alpha'^{-2}(\fft{\alpha'}{r})^{(7-p)}$ and relation between
Yang-Mills coupling and string coupling is given by
${g}_{YM}^2:=g_s (\alpha')^{(p-3)/2}$. We are interested in the
near-horizon limit by rescaling $r \to \alpha' r$ as $\alpha' \to
0$\footnote{Be aware that we have omit all numerical factors in
$H$ for its no relevance to our purpose. After recaling, r has the
dimension of mass instead of length.}. At this limit,
\begin{eqnarray}
&&ds^2=\alpha'(\fft{r^{(7-p)/2}}{\bar{g}_{YM}\sqrt{N}}d\vec{x}^2_{(p+1)}+
\fft{\bar{g}_{YM}\sqrt{N}}{r^{(7-p)/2}}dr^2+
\bar{g}_{YM}\sqrt{N}r^{(p-3)/2}d\Omega^2_{8-p})\nonumber\\
&&e^{\phi}=\bar{g}_{YM}^2 (\fft{\bar{g}_{YM}^2N}{r^{7-p}})^{(3-p)/4}
\end{eqnarray}
For simplicity, most of time we will simply put $H\to r^{p-7}$ by absorbing
those prefactors but recover them whenever
necessary. The ansatz becomes
\begin{eqnarray}\label{full_ansatz}
&&ds_{string}^2=
r^{-(p-7)/2}d\vec{x}^2_{(p+1)}+r^{(p-7)/2}(dr^2+r^2d\Omega^2_{(8-p)}), \\
&&e^\phi=r^{-(p-3)(p-7)/4}, \\
&&F_{0\cdots p r}=(7-p)r^{-(p-6)},
\end{eqnarray}

Sometimes it is convenient to work on the Einstein frame and dual
frame, which are obtained from weyl transformation such that
$g_{einstein}=e^{-\phi/2}g_{string}$ and
$g_{dual}=e^{2\phi/(p-7)}g_{string}$.  Their corresponding metric
become
\begin{equation}
ds_{einstein}^2=
r^{(p-7)^2/8}d\vec{x}^2_{(p+1)}+r^{(p-7)(p+1)/8}(dr^2+r^2d\Omega^2_{(8-p)})
\end{equation}
and
\begin{equation}
ds_{dual}^2= r^{-(p-5)}d\vec{x}^2_{(p+1)}+r^{-2}dr^2+ d\Omega^2_{(8-p)}
\end{equation}
The advantage of dual frame is that one can do another
transformation $u^2=R^2r^{(5-p)}$ to bring it to a AdS-like
coordinate such that
\begin{equation}\label{eqn:metric_ads}
ds_{AdS}^2=\fft{u^2}{R^2}d\vec{x}^2_{(p+1)}+R^2\fft{du^2}{u^2} +
d\Omega^2_{(8-p)}, \qquad R=2/(5-p)
\end{equation}
This near-horizon geometry is a domain-wall solution and is argued
to break $1/2$ of maximal supersymmetries and the quantum field
theory living on its boundary is discussed in \cite{BST}.

  In order to study the holographic RG flow for our new null-like domain wall solutions, we need the dimensionally reduced action for the following $(p+2)$-dimensional field configurations (in the dual frame)
\bea
&& ds^2=d\vec{y}^2_{(p+2)}+d\Omega^2_{(8-p)}\\
&& \phi=\phi(\vec{y})\\
&& F^2_{(p+2)}=-(p+2)! e^{2\gamma \phi}
\eea
and the resulting reduced action is
\be\label{dr}
S_{DR}=\int d^{p+2} y \sqrt{-g} e^{\gamma \phi} [R+{4(p-1)(p-4)\over (p-7)^2} (\partial \phi)^2 +{1\over 2} e^{(a-{(p-7)\gamma\over 4})\phi}]
\ee
where $\gamma=2(p-3)/(7-p)$. Note that we fix the 10-dimensional form flux and the $(8-p)$-sphere part in the metric in (null-like) domain wall solutions, but vary the $(p+2)$-dimensional metric and dilaton. In this way, the reduced action governs the dynamics for $\phi$ and $ds_{(p+2)}^2$.

Finally, we transform the action \eq{dr} to the one in Einstein
frame, the result is \be\label{draction} S_E=\int d^{p+2} y
\sqrt{-g} [R+{4(p-9)\over p(p-7)^2}(\partial \phi)^2 +{1\over 2}
e^{-{2\over p} \gamma \phi}]. \ee The field equations derived from
it can be solved by \bea\label{dtoe} && ds^2_{(p+2)}
=e^{{2\over p} \gamma \phi}[r^{5-p} a(u)^2 (-2dudv+h(u,r,\vec{x})du^2+d\vec{x}^2_{p-1})+r^{-2}dr^2]\\
&& e^{\phi}=r^{(3-p)(p-7)/4}a(u)^{(p+1)(p-7)/(2(p-3))}
\eea
with the functions  $a$ and $h$ given in section 3. Note that the overall Weyl factor in the metric is induced by the transformation from dual frame to Einstein one.

\section{Spin connections for null-like dilatonic domain-wall solutions}

Since the Killing spinor equations for type II supergravity are given in
string frame \cite{Berg}, we will translate the metric ansatz \eq{dual1} into
string frame and use the constraint \eq{cons1}, that is
\begin{equation}\label{eqn:metric_string}
ds_{string}^2=r^{2\alpha}a^{2\beta} (-2dudv+hdu^2+d\vec{x}_{(p-1)}^2)+r^{-2\alpha}a^{2\delta}(dr^2+r^2d\Omega_{(8-p)}^2)
\end{equation}
where
$\alpha=\fft{7-p}{4},\beta=\fft{p-7}{2(p-3)},\delta=-\fft{p+1}{2(p-3)}$.

It is convenient to choose a vielbein basis
\begin{eqnarray}\label{eqn:vielbein}
&&e^+=r^{\alpha}a^{\beta}du \nonumber\\
&&e^-=r^{\alpha}a^{\beta}(-dv+\fft12 h du) \nonumber\\
&&e^i=r^{\alpha}a^{\beta}dx^i, \nonumber\\
&&e^r=r^{-\alpha}a^{\delta}dr, \nonumber\\
&&e^m=r^{-\alpha}a^{\delta}d\Omega^m,
\end{eqnarray}

and work out the spin connections
\begin{eqnarray}
&&\omega^+{}_r=\alpha r^{\alpha-1} a^{-\delta}e^+,\nonumber\\
&&\omega^-{}_r=\alpha r^{\alpha-1} a^{-\delta}e^- + \fft12
r^{\alpha} a^{-\delta}\partial_rh e^+, \nonumber\\
&&\omega^-{}_+=2 \beta r^{-\alpha} a^{-\beta-1}\partial_u a
e^-,\nonumber\\
&&\omega^-{}_i=\fft12\partial_ih r^{-\alpha} a^{-\beta}
e^+,\nonumber\\
&&\omega^i{}_r=\alpha r^{\alpha-1} a^{-\delta}e^i, \qquad
\omega^i{}_+=2\beta r^{-\alpha} a^{-\beta-1} \partial_u a e^i,
\nonumber\\
&&\omega^r{}_+=2\delta r^{-\alpha} a^{-\beta-1}\partial_u a
e^r,\nonumber\\
&&\omega^m{}_r=(-\alpha+1)r^{\alpha-1} a^{-\delta} e^m, \qquad
\omega^m{}_+=2\delta r^{-\alpha} a^{-\beta-1} \partial_u a e^m
\label{eqn:connection}
\end{eqnarray}
as well as the gamma matrices in vielbein basis,
\begin{eqnarray}
&&\bar{\gamma}^+=r^{\alpha}a^{\beta}\gamma^u, \qquad
\bar{\gamma}^-=-r^{\alpha}a^{\beta}\gamma^v + \fft12 h r^{\alpha}a^{\beta}\gamma^u,\nonumber\\
&&\bar{\gamma}^r=r^{-\alpha}a^{\delta}\gamma^r, \qquad
\bar{\gamma}^i=r^{\alpha}a^{\beta}\gamma^i, \qquad \bar{\gamma}^m=r^{-\alpha}a^{\delta}\gamma^{\Omega^{m}} \nonumber\\
\end{eqnarray}

\section{$AdS_5\times S^5$ in the null-like axi-dilaton background}
For completeness, we reproduce the solutions found in
\cite{ChuHo}, where near-horizon geometry is $AdS_5\times S^5$
with null-like axi-dilaton background. Starting with the ansatz
\begin{eqnarray}
&&ds^2=e^{2\rho}a(u)^2(-2dudv+hdu^2+d\vec{x}^2_{(2)})+d\rho^2+d\Omega^2_{(5)}\nonumber\\
&&e^\phi=e^{\phi(u)}, \nonumber\\
&&F_{(5)}=4(e^{4\rho}a(u)^4du\wedge dv\wedge\cdots\wedge
d\rho+ \omega_{(5)}).
\end{eqnarray}
Here $\omega_{(5)}$ is the volume form of $5$-sphere.  Notice that
the five-form field strength is self-dual such that $F^2_{(5)}=0$.

The dilaton field is decoupled at $p=3$ ($\sigma=0$), thus equation of
motion becomes
\begin{eqnarray}
&&R_{MN}=\fft12\partial_M \phi \partial_N \phi + \fft12
e^{2\phi}\partial_M\chi\partial_N\chi + \fft{1}{48}
(F_{MK_1\cdots K_{4}}F_N^{\;\;K_1\cdots K_{4}}-\fft{1}{10}g_{MN}F_{(5)}^2), \label{ads5_eom_einstein}\\
&&\nabla_M ( F^{MN_1\cdots N_{4}})=0,  \\
&&\nabla^M\nabla_M \phi = e^{2\phi}\partial^{M}\partial_M\chi,
\label{ads5_eom_phi} \qquad  \nabla_M(e^{2\phi}\nabla^M\chi)=0,
\end{eqnarray}
where we have included the axion field as well.

Ricci tensors are calculated,
\begin{eqnarray}
&&R_{uu}=2((\partial_u \ln a)^2-\partial^2_u \ln a)-\fft12 \vec{\nabla}^2h-e^{2\rho}a^2(4h+2\partial_{\rho}h+\fft12\partial_{\rho}^2h)\\
&&R_{uv}=4e^{2\rho}a^2\\
&&R_{ij}=-4e^{2\rho}a^2\delta_{ij}\\
&&R_{\rho\rho}=-2\\
&&R_{mn}=4g^{(5-sphere)}_{mn}
\end{eqnarray}
where $g^{(5-sphere)}_{mn}$ is the metric for unit 5-sphere.

Equation \eq{ads5_eom_phi} is automatically satisfied due to
$g^{uu}=0$. Then only the $uu$-component of equation of motion
gives the nontrivial constraint,
\begin{equation}\label{ads5_eom_uu}
2((\partial_u \ln a)^2-\partial^2_u \ln
a)-\fft12\vec{\nabla}^2 h-e^{2\rho}a^2
(2\partial_{\rho}h+\fft12\partial_{\rho}^2h)=\fft12\partial_u\phi\partial_u\phi+\fft12e^{2\phi}\partial_u\chi\partial_u\chi.
\end{equation}
Although our ansatz starts with vanishing axion field, it is
straightforward to generate a class of solutions with both
u-dependent axi-dilaton switched on via the well-known S-duality
transformation:
\begin{eqnarray}\label{eqn:s-dual}
&&\tau \to \fft{a\tau+b}{c\tau+d}, \qquad
H^{\alpha} \to (\Lambda^T)^{-1}{}^{\alpha}{}_{\beta}H^{\beta}, \nonumber\\
&& \tau=\chi+ie^{-\phi}, \qquad \Lambda = \left( \begin{array}{cc} a&b\\
c&d\end{array}\right) \in SL(2,R), \qquad H^{\alpha}=\left(
\begin{array}{c}H^{NSNS}_{(3)}\\ H^{RR}_{(3)}\end{array}\right),
\end{eqnarray}
here the transformation of three-form flux is redundant for
vanishing $H_{(3)}$.

\section{$AdS_3 \times S^3\times M_4$ in the null-like axi-dilaton background}
For completeness, we also give time dependent solution for the
$AdS_3 \times S^3\times M_4$ vaccum.  The relevant equations of
motion are
\begin{eqnarray}
&&R_{MN}=\fft12 \partial_{M}\phi\partial_{N}\phi+ \fft12 e^{2\phi}\partial_{M}\chi\partial_{N}\chi + \fft14 e^{-\phi}H_{MPQ}H_{N}{}^{PQ}+\fft14 e^{\phi}\tilde{H}_{MPQ}\tilde{H}_N{}^{PQ},\\
&&\nabla^M\nabla_M\phi+e^{2\phi}\partial^M\partial_M\chi + \fft{1}{12} (e^{-\phi}H^2-e^{\phi}\tilde{H}^2)=0, \label{eq:ads3_eom_chi}\\
&&\nabla_{M}(\sqrt{-g}e^{-\phi}H^{MNP})=\nabla_M
(\sqrt{-g}e^{\phi}\chi\tilde{H}^{MNP}), \qquad  \nabla_{M}(\sqrt{-g}e^{\phi}\tilde{H}^{MNP})=0,\\
&&\nabla_{M}(e^{2\phi}\nabla^{M}\chi)+\fft16\tilde{H}_{MNP}H^{MNP}=0,
\end{eqnarray}
where $H_{(3)}\equiv H^{NSNS}$ and $\tilde{H}_{(3)}\equiv
H^{RR}-\chi H^{NSNS}$.  They can be solved by following ansatz:
\begin{eqnarray}
&&ds^2=e^{2\rho}a(u)^2(-2dudv+h(u,\rho)du^2)+d\rho^2+d\Omega_3^2+d\vec{x}_i^2,\nonumber\\
&&H=  2(e^{2\rho}a(u)^2 du\wedge dv\wedge d\rho + \omega_{(3)}),\nonumber\\
&&\tilde{H}=0, \nonumber\\
&&\chi=\chi(u),
\end{eqnarray}
where $x_i$'s are coordinates four dimensional flat space.  Since
$H$ is self-dual in the six dimensions $AdS_3\times S^3$,
\eq{eq:ads3_eom_chi} is satisfied.  Then the non-vanishing Ricci
tensors are calculated:
\begin{eqnarray}
&&R_{uu}=-\fft12 e^{2\rho}a^2(4h+2\partial_{\rho} h+\partial^2_{\rho} h),\\
&&R_{uv}=2 e^{2\rho}a^2\\
&&R_{\rho\rho}=-2, \\
&&R_{mn}=2g^{(3-sphere)}_{mn}.
\end{eqnarray}

The $uu$-component again gives the nontrivial constraint:
\begin{equation}
-e^{2\rho}a^2(2\partial_{\rho} h+\partial^2_{\rho}
h)=e^{2\phi}\partial_u\chi\partial_u\chi+\partial_u\phi\partial_u\phi,
\end{equation}
Since the RHS only allows u-dependence, this is solved by
\begin{equation}
\partial_{\rho}^2h+2\partial_{\rho}h = P(u)e^{-2\rho},
\end{equation}
provided arbitrary profile $P(u)$.  Again, even though we start
with ansatz of vanishing scalar, it is easy to construct a class
of solution with both axi-dilaton turned on via the transformation
given by \eq{eqn:s-dual}.

\end{appendix}
%%%%%%%%%%%%%%%%%%%%%%%%%%%%%%%%%%%%%%%%%%%%%%%%%%%%%%%%%%%%%%%%%%%%%%%%%%%%

\end{document}